\documentclass[a4paper,10pt]{article}
\usepackage{qite}
\usepackage{amsmath,amssymb}
\usepackage{mymacro,graphicx}
\usepackage{txfonts}
\begin{document}
\title{
 Proposal of an eavesdropping experiment for BB84 QKD protocol with 1$\rightarrow$3 phase-covariant quantum cloner
}
\author{
  Yuta Okubo\makebox[0pt][l]{\addressmark{1}} \\
  yokubo@qci.jst.go.jp
  \and
  Francesco Buscemi\makebox[0pt][l]{\addressmark{2}}\\
  buscemi@qci.jst.go.jp
  \and
  Akihisa Tomita\makebox[0pt][l]{\addressmark{2}}\\
  tomita@qci.jst.go.jp
}
\seteaddress{
\makebox[0pt][r]{\addressmark{1}}%
Department of Frontier Science, University of Tsukuba\\
1-1-1, Tennodai, Tsukuba, Ibaraki, Japan\\
TEL: +81-29-850-1110+2339 \quad FAX: +81-29-856-6139\\
\makebox[0pt][r]{\addressmark{2}}%
ERATO-SORST Quantum Computation and Information Project, Japan Science and Technology Agency\\ 5-28-3, Hongo, Bunkyo-ku, Tokyo, Japan\\
}
\abstract{ 
We propose an eavesdropping experiment with linear optical 1$\rightarrow$3 phase-covariant quantum cloner. In this paper, we have designed an optical circuit of the cloner and shown how the eavesdropper (Eve) utilizes her clones. We have also optimized the measurement scheme for Eve by numerical calculation. The optimized measurement is easy to implement with liner optics.
}
\maketitle
\section{Introduction}
Quantum key distribution, which ensures the secure communication between two parties, is one of the most promising technologies in the field of quantum information processing.
Many researches about the security of QKD have been done not only in the ideal condition \cite{shor} but  practical conditions\cite{GLLP}.
However, there have been few experimental research of such security analysis i.e. attempt of eavesdropping.
In this paper we explore an eavesdropping experiment with a quantum cloner using linear optics, because it is well known that linear optics is the useful tool to construct small-scale quantum computers.

\section{Phase-covariant quantum cloning}
Although it is impossible to prepare the perfect clone of quantum states (No-cloning theorem), imperfect cloning can be achieved. 
Firstly Buzek and Hillery\cite{buz} proposed the universal quantum cloner, which duplicates the unknown qubits.
However the universal cloning is superfluous for eavesdropping BB84 protocol, since the input state is restricted to the part of Hilbert space.
 On the other hand, phase-covariant cloning\cite{pcm} offers a better quality of output clones, at the cost of the restriction for the input state to the equator of the Bloch sphere.
Hence we utilize phase-covariant cloning for eavesdropping.
\subsection{Experimental setup}
\begin{figure}[ht]
\begin{center}
\includegraphics[width=.99\linewidth]{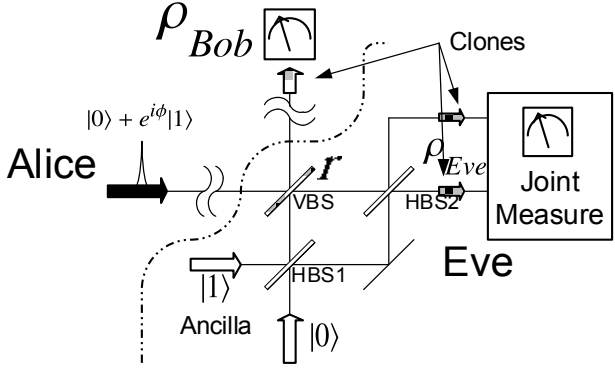}
\caption{Schematic of the eavesdropping circuit with linear optics. HBS: Half beam splitter, VBS: Variable ratio beam splitter}\label{exp}
\end{center}
\end{figure}

An experimental setup to implement linear optical 1$\rightarrow$3 phase-covariant quantum cloner\cite{Origin} is shown in Fig.\ref{exp}
This setup consists of the beam splitter array, called $tritter$\cite{tritter}, one of the three inputs is for Alice qubit and the others are for Eve's ancillas. 
Here we assume classical bit "0", "1" of Alice is encoded to 
$\Bigl\{\ket{+x}=\frac{\ket{0}+\ket{1}}{\sqrt{2}},\ket{-x}=\frac{\ket{0}-\ket{1}}{\sqrt{2}}\Bigr\}$ or $\Bigl\{\ket{+y}=\frac{\ket{0}+i\ket{1}}{\sqrt{2}},\ket{-y}=\frac{\ket{0}-i\ket{1}}{\sqrt{2}}\Bigr\}$, respectively.
Hence the input state is written by
\begin{align}
\ket{\Psi_\text{in}}\equiv&\frac{\ket{0}+e^{i\phi}\ket{1}}{\sqrt{2}}\\
\phi=&\{0,\pi,-\tfrac{\pi}{2},\tfrac{\pi}{2}\}\notag
\end{align}
and these are physically represented by the photonic polarization or the  relative phase between two coherent pulses (time-bin).
After combining the three inputs\footnote{We $post$-$selected$ the events where one and only one photon is detected on one detector.} at the tritter, the tripartite entangled state $\ket{\xi(\phi,r)}$ is obtained as follows;
\begin{align}
\ket{&\Psi_\text{in}}\otimes\ket{0}\ket{1}_\text{ancilla}\rightarrow\ket{\xi(\phi,r)}\notag\\
&=\frac{i(r-1)\sqrt{r}}{2}
\ket{0}_\text{Bob}\ket{0}_\text{Eve$_1$}\ket{1}_\text{Eve$_2$}\notag\\
&\hspace{15pt}+\frac{1}{4}(2ir^{3/2}+3r-2i\sqrt{r}-1)
\ket{0}_\text{Bob}\ket{1}_\text{Eve$_1$}\ket{0}_\text{Eve$_2$}\notag\\
&\hspace{35pt}+\frac{1}{4}(2ir^{3/2}-3r-2i\sqrt{r}+1)
\ket{1}_\text{Bob}\ket{0}_\text{Eve$_1$}\ket{0}_\text{Eve$_2$}\notag\\
&\hspace{10pt}+e^{i\phi}\Biggl(\ \frac{i(r-1)\sqrt{r}}{2}
\ket{1}_\text{Bob}\ket{1}_\text{Eve$_1$}\ket{0}_\text{Eve$_2$}\notag\\
&\hspace{40pt}+\frac{1}{4}(2ir^{3/2}-3r-2i\sqrt{r}+1)
\ket{1}_\text{Bob}\ket{0}_\text{Eve$_1$}\ket{1}_\text{Eve$_2$}\notag\\
&\hspace{60pt}+\frac{1}{4}(2ir^{3/2}+3r-2i\sqrt{r}-1)
\ket{0}_\text{Bob}\ket{1}_\text{Eve$_1$}\ket{1}_\text{Eve$_2$}\Biggr)
\end{align}
where $r$ is the branching ratio of the variable ratio beam splitter (VBS).
Eve sends the first qubit$\ \ket{}_\text{Bob}$ to Bob and keeps the remaining two qubits$\ \ket{}_\text{Eve$_1$}\ket{}_\text{Eve$_2$}$.
The success probability of this cloning operation is
\begin{align}
P_{suc}(r)=\frac{1-3r^2+6r^3}{4}.
\end{align}
We plot it in Fig.\ref{fidelity}
\subsection{Single qubit fidelities}
\begin{figure}[ht]
\begin{center}
\includegraphics[width=.9\linewidth]{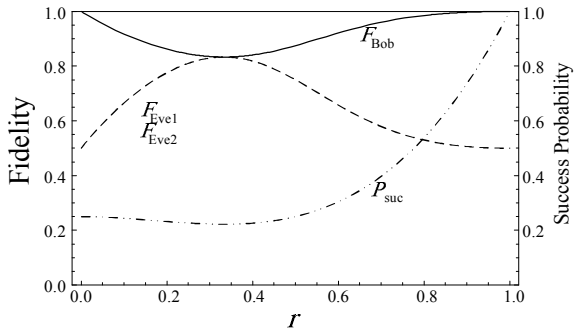}
\caption{Single qubit fidelity v.s. $r$ (branching ratio of BS) and success probability}
\label{fidelity}
\end{center}
\end{figure}

The single qubit fidelities, which indicate the performance of the cloner, can be calculated as follows
\begin{align}
F_\text{Bob}(r)&=\bra{\Psi_\text{in}}\rho_\text{Bob}\ket{\Psi_\text{in}}\notag\\
&=-\frac{-1+r+r^2-5r^3}{1-3r^2+6r^3},\label{444}\\
F_\text{Eve1}(r)&=\bra{\Psi_\text{in}}\rho_\text{Eve1}\ket{\Psi_\text{in}}\notag\\
&=\frac{1+4r-11r^2+10r^3}{2-6r^2+12r^3},\\
F_\text{Eve2}(r)&=\bra{\Psi_\text{in}}\rho_\text{Eve2}\ket{\Psi_\text{in}}\notag\\
&=\frac{1+4r-11r^2+10r^3}{2-6r^2+12r^3}.
\end{align}
where $\rho_\text{Bob}, \rho_\text{Eve1}, \rho_\text{Eve2}$ are local density matrices obtained by partial-tracing the whole density matrix $\ketbra{\xi}$ over the other qubits.
We plot $F_\text{Bob}$, $F_\text{Eve1}$ and $F_\text{Eve2}$ as a function of the branching ratio of $r$ in Fig.\ref{fidelity}.
It should be noted that the clone for Bob is perfect, when $r=0$, or $r=1$. In both case, Eve has no information about the input state.
The first case is trivial, since the input state is perfectly reflected by the VBS. On the other hand, the latter case corresponds to the quantum teleportation, where the the input state has been teleported to Bob's side.\footnote{Because of the $post$-$selection$, the entangled state has been generated in HBS1. Then, conditional Bell state measurement\cite{tele} has been achieved by HBS2 }
\section{Eavesdropping}
In order to evaluate the performance of the proposed cloner as an eavesdropper, we derive the relation between error rate of Bob and the amount of information obtained by Eve.
\subsection{Bit errors}
The quantum bit error rate (QBER) is written by
\begin{align}
\text{QBER}=&\frac{(1-p_B(0))(1-F_\text{Bob})+p_B(0)p_d}{1-p_B(0)+2p_B(0)p_d}\label{qber}\\
\xrightarrow{p_d=0}&1-F_\text{Bob}\equiv D.\notag\\
&p_d : \text{ dark count}\notag\\
&p_B(0): \text{Probability that Bob detects no photon.}\notag
\end{align}
where $D$ is disturbance that characterizes the intrinsic error of communication channel,
and Eve can control it by changing the $r$(branching ratio) as shown in Eq.\eqref{444}.
As shown in Eq.\eqref{qber}, QBER is equivalent to $D$, if there is no dark count on Bob.
Since Eve cannot draw any information from the dark count contribution, we focus on disturbance in the rest of this paper.
\subsection{Eve's information}
Let us calculate the amount of Eve's information $I_\text{Eve}$ as a function of disturbance.
We assume that Eve can keep her qubits on quantum memory until Alice reveals the basis. Eve knows the basis on which Alice encoded her classical bit, so she only has to discriminate ${\rho_\text{Eve}(0), \rho_\text{Eve}(\pi)}$ $\bigl( \text{or}\ \rho_\text{Eve}$ $(-\pi/2)$,$\rho_\text{Eve}(\pi/2)\bigr)$ to obtain the information about the input state, where $\rho_\text{Eve}(\phi)$ denotes Eve's local density matrix defined as follows;
\begin{align}
\rho_\text{Eve}\equiv& \Tr_\text{Bob}\ketbra{\xi(\phi,r)}.
\end{align}
Eve's information $I_\text{Eve}$is written by
\begin{align}
I_\text{Eve}(\hat{M}_k)=&\sum_k {\Tr} \hat{R}_k\hat{M}_k\\
\hat{R}_k\equiv&\sum_j \log
\frac{\Tr(\rho_j \hat{M}_k)}
{\sum_l\Tr \rho_j \hat{M}_l\sum_m\Tr(\rho_m \hat{M}_k)}.\\
\{\rho_j\}\equiv&
\begin{cases}
 \{\rho_\text{Eve}(0),\rho_\text{Eve}(\pi)\}\ \text{for $x$-basis}\\
\{\rho_\text{Eve}(-\pi/2),\rho_\text{Eve}(\pi/2)\}\ \text{for $y$-basis}
\end{cases}
\end{align}
where $\hat{M}_k$ is Eve's measurement operator.
This implies that the amount of information refers to how well Eve can discriminate the state by choosing appropriate measurement of $\{\hat{M}_k\}$. 
We calculated $I_\text{Eve}$ with the following conventional measurement and plotted it in Fig.\ref{inf}.
\begin{align}
\begin{cases}
\hat{M}_1^0=\ketbra{+}\otimes\ketbra{+}\\
\hat{M}_2^0=\ketbra{+}\otimes\ketbra{-}\\
\hat{M}_3^0=\ketbra{-}\otimes\ketbra{+}\\
\hat{M}_4^0=\ketbra{-}\otimes\ketbra{-}
\end{cases}
\end{align}
Note that these measurements are same as Bob's.
\subsection{Optimal POVM}
\begin{figure}[ht]
\begin{center}
\includegraphics[width=1\linewidth]{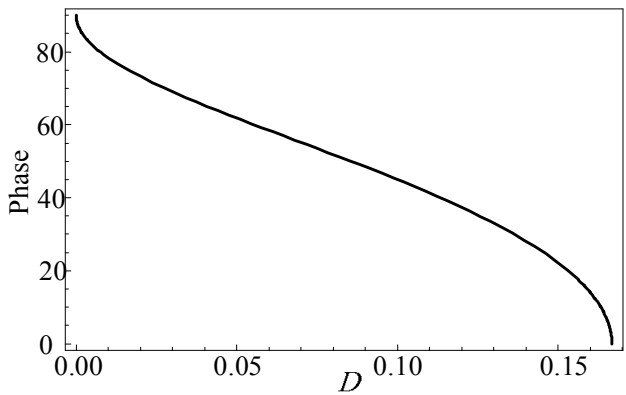}
\caption{The phase factor $\theta$ in the optimal POVM}\label{theta}
\end{center}
\end{figure}

\begin{figure}[ht]
\begin{center}
\includegraphics[width=1\linewidth]{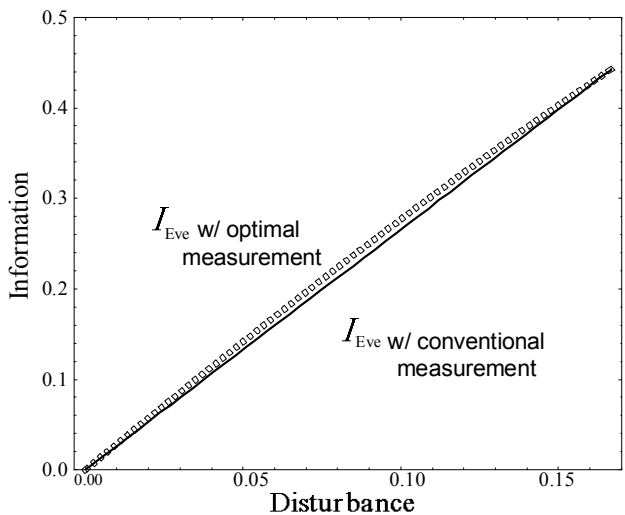}
\caption{Disturbance v.s. Information}\label{inf}
\end{center}
\end{figure}

We have optimized POVMs by the numerical calculation\cite{iter}
The optimized POVMs are given by
\begin{align}
\begin{cases}
\hat{M}_1= \ketbra{\chi_{-\theta}}\otimes\ketbra{\chi_{\theta}}\\
\hat{M}_2= \ketbra{\chi_{-\theta}}\otimes\ketbra{\chi_{\pi+\theta}}\\
\hat{M}_3= \ketbra{\chi_{\pi-\theta}}\otimes\ketbra{\chi_{\theta}}\\
\hat{M}_4= \ketbra{\chi_{\pi-\theta}}\otimes\ketbra{\chi_{\pi+\theta}}.
\end{cases}\\
\hspace{10pt}\ket{\chi_\theta}\equiv\frac{\ket{0}+e^{i\theta}\ket{1}}{\sqrt{2}},
\hspace{10pt}\cos\theta=\tfrac{2\sqrt{D}}{\sqrt{1-2D}}\notag
\end{align}
The phase factor $\theta$ is determined by  the disturbance, as plotted in Fig.\ref{theta}.
We plot $I_\text{Eve}$ with the optimal POVMs in Fig.\ref{inf}, which corresponds to the accessible information\cite{Iac}.
The optimal measurement makes an improvement over conventional measurement.
It should be noted that these optimal POVMs are separable. Moreover it is easy to implement with linear optics, because the optimal POVMS are realized by adjusting the measurement basis (i.e. the angle of the polarizers) for each output port.
\newpage
In conclusion, we proposed an eavesdropping experiment for BB84 protocol.
We have designed not only the optical cloning circuit but the measurement scheme in order that Eve obtains the accessible information.

\end{document}